\documentclass[12pt,aps,prb,preprint,amsmath,amssymb]{revtex4}   

\usepackage{amsmath}    
\usepackage{graphicx}   
\usepackage{color}

\begin{document}

\title{From permanence to total availability: a quantum conceptual upgrade}
\author{Massimiliano Sassoli de Bianchi}
\affiliation{Laboratorio di Autoricerca di Base, 6914 Carona, Switzerland}\date{\today}
\email{autoricerca@gmail.com}   

\begin{abstract}
\noindent
We consider the classical concept of \textit{time of permanence} and observe that its quantum equivalent is described by a bona fide self-adjoint operator. Its interpretation, by means of the spectral theorem, reveals that we have to abandon not only the idea that quantum entities would be characterizable in terms of spatial trajectories but, more generally, that they would possess the very attribute of \textit{spatiality}. Consequently, a \textit{permanence time} shouldn't be interpreted as a ``time'' in quantum mechanics, but as a measure of the \textit{total availability} of a quantum entity in participating to a  process of creation of a spatial localization.   
\end{abstract}

\maketitle

Most probably, the creation of the very concept of time has been historically motivated by the observation of entities moving in \textit{space}, along \textit{trajectories} (think for example to our sun, as our first rudimental clock). Hence, we can assume that time is primarily a classical concept, pertaining to the realm of the macro-objects, moving in our three-dimensional Euclidean space. 

On the other hand, contrary to classical entities, quantum entities are notoriously recalcitrant in letting themselves being described in terms of trajectories. It is then natural to ask if, at the quantum level of reality, time would remain a meaningful concept and, if not, to inquire about what would be a possible new, upgraded concept, constituting its natural generalization. These are the fundamental questions we are going to tentatively explore in the present article.

To do this, we shall proceed as follows. In Sec.~\ref{Time of arrival}, we start by illustrating what is a well known difficulty in quantum mechanics (QM): the absence of a self-adjoint \textit{arrival time operator}. In Sec.~\ref{Classical time of permanence}, we observe that in classical mechanics (CM) we can always replace, without losing generality, the concept of arrival time by the concept of \textit{time of permanence}, and that the latter can be equivalently defined as a time-integral over probabilities of presence. 

Then, considering that the probability of presence is a well defined concept in QM, in Sec.~\ref{Quantum time of permanence} we show how to define a meaningful quantum time of permanence, as the average of a bona-fide self-adjoint operator having the good property of commuting with the energy operator. In Sec.~\ref{An explicit calculation}, we explicitly calculate the quantum time of permanence for the simple situation of a one-dimensional free evolving quantum entity with positive momentum, showing that it coincides with the classical time, in the monoenergetic limit. 

In Sec.~\ref{Interferences}, we go on by explaining how the pure quantum effect of interference (which occurs in QM as a consequence of the non compatibility of certain observables) manifests at the level of the permanence time. In Sec.~\ref{Non-spatiality}, we exploit a mathematical result proved by Jaworski~\cite{Jaw} to show that the very notion of permanence time forces us to fully abandon the classical notion of spatial trajectory for quantum entities and, consequently, the false prejudice that they would be spatial entities, as already suggested by Diederik Aerts in a number of works (see for instance Refs.~\onlinecite{Aer, Aer2}) and also recently emphasized  by the present author in this journal~\cite{Sas3}. 

This lack of spatiality of quantum entities implies that the very concept of time of presence (in a given region of space) needs to be revisited in QM. This we shall do in Sec.~\ref{Total availability}, exploiting the conceptual framework of the so-called \textit{creation-discovery view}, that was developed by Aerts and his group in Brussels~\cite{Aer, Aer2, Aer3}, and particularly the key notion of \textit{availability}, which follows intrinsically from it. (For the commodity of the reader, some of the important ideas of the creation-discovery view will be summarized in Sec.~\ref{The creation-discovery view}).

More precisely, by proposing to understand quantum probabilities as a measure of the \textit{degree of availability} of quantum entities, in lending themselves to specific experiences (like the one of creating a spatial localization), we show that the classical concept of time of presence should be upgraded in QM to the more general concept of \textit{total availability}, which has the advantage of remaining fully consistent also for non-spatial entities.   

Based on the discussion of Sec.~\ref{Total availability}, we then pursue our conceptual analysis in Sec.~\ref{Availability change}, proposing to reinterpret the well-known concept of \textit{global time-delay} as a \textit{total availability shift} experienced by a non-spatial quantum entity, as a consequence of the ``switching on'' of the interaction. Finally, in Sec.~\ref{discussion}, we also consider the concept of total availability in a relational sense, and provide some concluding remarks.

\section{Time of arrival}\label{Time of arrival}

We consider a classical free particle of mass $m$, and denote by $\textbf{q}(t)$ and $\textbf{p}(t)$ its position and momentum, respectively (three-vectors are in bold type). Being the particle free, its momentum is conserved, i.e., $\textbf{p}(t)=\textbf{p}$, and $\textbf{q}(t)= \textbf{q} + \textbf{p}t/m$, where $\textbf{q}$ and $\textbf{p}$ are the particle's position and momentum at time $t=0$, respectively. Setting $\hat{\textbf{p}} =\textbf{p} /|\textbf{p}|$, and defining the particle's scalar velocity $v=|\textbf{p}|/m=\sqrt{2E/m}$ ($E$ is the kinetic energy), we can write
\begin{equation}
\label{particle equation}
t= \frac{1}{v}\left[\textbf{q}(t)\hat{\textbf{p}} - \hat{\textbf{p}}\textbf{q}\right].
\end{equation}
Clearly, if we set $\textbf{q}(t)=0$ in (\ref{particle equation}), we obtain that
\begin{equation}
\label{arrival time at the origin}
t= -\frac{1}{v}\hat{\textbf{p}}\textbf{q}=-\frac{1}{2E}\textbf{p}\textbf{q}
\end{equation}
is the arrival time of the particle at the plan passing from the origin, orthogonal to the direction of  movement. 

From the classical observable (\ref{arrival time at the origin}), we can try to construct the corresponding arrival time quantum observable. For this, we apply the standard quantization rule that consists in replacing in a classical expression the position and momentum variables by the corresponding position and momentum operators (we use capital letters to distinguish quantum operators from classical variables) $\textbf{q}\rightarrow\textbf{Q}$, $\textbf{p}\rightarrow\textbf{P}$, then symmetrizing all products of non-commuting observables. If we do so, we obtain the following candidate for an arrival time observable: 
\begin{equation}
\label{time operator}
T_0=-\frac{1}{4}\left(H_0^{-1}\textbf{P}\textbf{Q}+\textbf{Q}\textbf{P} H_0^{-1}\right),
\end{equation}
where $H_0 = \textbf{P}^2/2m$ is the free Hamiltonian. 

The above formal procedure is however not without difficulties. Indeed, the arrival time operator $T_0$ is not self-adjoint. This can be easily shown using a famous argument due to Pauli (see for instance the discussion in Ref.~\onlinecite{Hilgevoord}): from the canonical commutation relations between position and momentum, one can easily show that $T_0$ obeys the canonical commutation relation with the free Hamiltonian: $\left[H_0,T_0\right]=i\hbar \mathbb{I}$. Hence, if it would be self-adjoint, then $\exp\left(i\alpha T_0\right)$ would be a unitary representation of the group of energy translations, and as we can translate both to the right and to the left, the very existence of such a representation would be in contradiction with the boundedness from below of the spectrum of $H_0$. Thus, $T_0$ cannot be self-adjoint and $H_0$ doesn't possess a canonically conjugate operator.

To understand why self-adjointness is an important requisite, we recall that in physics a system is described in terms of its properties, and that in standard QM properties correspond to orthogonal projectors, whose expectation values over the state of the system give the \textit{a priori} probabilities for properties being confirmed by experiments, and this independently of the specificities of the measuring apparatus. The request to represent physical observables by (densely defined) self-adjoint operators then follows from the spectral theorem, which allows to \textit{uniquely} decompose a self-adjoint operator by means of a projection-valued measure, and therefore to unambiguously relate the measure of the observable to the properties of the system~\footnote{The difficulty of not disposing of self-adjoint operators for arrival time observables is usually believed to be related to the essentially different role that time would play in quantum physics in comparison to classical physics. However, as it has been lucidly pointed out by Hilgevoord~\cite{Hilgevoord}, the problem is only apparent and results from a confusion between the time coordinate (the partner of the space coordinate of the space-time reference frame), which needs not be quantized, and the time variables, which are ordinary dynamical variables, measured by specific instruments, called \textit{clocks}. Of course, QM being not classical mechanics (CM), one must be prepared to encounter situations where quantum dynamical time variables cannot always be defined as in the corresponding classical situation. This is exactly what happens when dealing with arrival times, which cannot be defined in terms of self-adjoint operators, but only in terms of symmetric operators. This means that one has to renounce to decompose the time operator in terms of projection-valued measures, using instead more general positive operator-valued measures. The price to be paid is that then the projection postulate no longer holds and arrival time observables cannot anymore be uniquely defined and will in general depend on the detailed description of the experimental apparatus used to carry out the measure.}.

\section{Classical time of permanence}\label{Classical time of permanence}

Coming back to our classical particle, we can observe that, as it is the case with positions, we never measure time instants in absolute terms, but always in relative terms, i.e., we measure \textit{time intervals}, or \textit{durations}. 

When for instance we tell somebody that a train will arrive at the railway station at, say, 16:00, what we mean is that a time interval of 16 hours has elapsed between the following two events: ``our watch indicates 00:00'' and ``the train arrives at the railway station.'' The reason why we usually forget to mention the first event is that we assume that our interlocutor's clock is duly synchronized with our, so that we share a same time origin. If, on the contrary, we suspect this not being the case, then we certainly need to make a more precise statement, making explicit our time zone, or giving whatever other relevant information that would allow the other observer to unambiguously determine the time origin with respect to which we have measured the train's arrival time.

In the same way, when we say that a classical evolving particle arrives, say, at a distance $r$ from the origin at time $t(r)$, what we truly mean, more exactly, is that a \textit{time interval} $\Delta t(r) =t(r)-0$ has elapsed between the following two events: ``the laboratory clock indicates 00:00'' and ``the particle arrives at a distance $r$ from the origin.'' But of course, the choice of the time origin of the laboratory clock is completely arbitrary, and one is free to change it according to one's preferences. 

Let us assume, for simplicity, that the particle moves freely. If we take $r$ large enough, we know that it will arrive at a distance $r$ from the origin exactly twice during its evolution. Let us call $t_-(r)$ and $t_+(r)$ these two arrival times, respectively. As we said, what we truly mean by them is the time intervals $\Delta t_-(r) =t_-(r)-0$, and $\Delta t_+(r) =t_+(r)-0$.

Now, as we are free to do it, let us choose to set the zero of the laboratory's clock in coincidence with the first time at which the particle arrives at the distance $r$ from the origin. This amounts to consider a new inertial frame, specified by the following time-shift transformation: $t\rightarrow t-t_-(r)$. In this new inertial frame, the arrival time $t_+(r)$ becomes $t_+(r)\rightarrow t_+(r)-t_-(r)$. In other terms, in a frame of reference having the time origin at $t_-(r)$, the arrival time $t_+(r)$ becomes equal to (the superscript ``$0$'' indicates that the particle is free evolving)
\begin{equation}
\label{sojourn time classical}
T^0(B_r)= t_+(r)-t_-(r),
\end{equation}
which is the time spent by the particle inside the ball $B_r$ of radius $r$, centered at the origin of the spatial system of coordinates. Let us call this duration the \textit{time of permanence} (or of \textit{sojourn}) of the classical particle inside the spatial region $B_r$. 

Following our above simple analysis, we observe that in classical mechanics (CM) we can always replace, without losing generality, the concept of arrival time by the concept of time of permanence inside a given region of space. It is therefore natural to ask if it wouldn't be the classical concept of time of permanence, instead of the one of time of arrival, that should be consistently transmigrated into the quantum realm. Of course, one may object that if there is no place for arrival times in QM, as self-adjoint observables, then the same must be true for notions which are classically related to them, like for instance times of permanence which, as we have seen, are defined as a difference (or sum of differences) of arrival times. 

To see why this is not the case, we start by observing that it is perfectly possible to make sense in CM of a notion of time of permanence without making any explicit reference to a notion of arrival time. For this, let us adopt a probabilistic perspective and consider the probability ${\cal P}_t(B_r)$ for the particle to be present inside $B_r$ at time $t$. Integrating this probability over all time instants, we can calculate the average time the particle spends in total inside $B_r$ by: 
\begin{equation}
\label{sojourn-time-definition}
T^0(B_r) = \int_{-\infty}^{\infty}dt\, {\cal P}_t(B_r).
\end{equation} 

Defining the time of permanence of the particle as a sum over probabilities of presence is a natural procedure if one only possesses a statistical knowledge of the particle's trajectory, as it is the case for instance when its initial condition is described by a probability distribution in phase space. However, definition (\ref{sojourn-time-definition}) makes  full sense also when the particle's dynamics is perfectly known. Hence, it constitutes an alternative, more general definition for the time of permanence, which is equivalent to the one obtained in terms of a difference of arrival times, when a notion of trajectory is available. 

Let us show this more explicitly, and for sake of simplicity let us limit ourselves to the one-dimensional case. Then, the one-dimensional ball $B_r$ of radius $r$, centered at the origin, reduces to the interval $[-r,r]$. Assuming the particle's momentum $p>0$ (it moves from the left to the right), we can easily solve (\ref{particle equation}), setting $|q(t)|=r$, and find that $t_+(r)=\frac{1}{v}(r-q)$ and $t_-(r)=-\frac{1}{v}(r+q)$, so that their difference (\ref{sojourn time classical}) is $T^0(B_r)=2r/v$. 

On the other hand, since the particle's probability of being present inside $B_r$ is equal to $1$, if $q(t)\in [-r,r]$, and zero otherwise, we have ${\cal P}_t(B_r)=\chi_r[q(t)]$, where $\chi_r(x)$ is the characteristic function of the interval $[-r,r]$. Thus, the time of permanence (\ref{sojourn-time-definition}) is given by:
\begin{equation}
\label{sojourn-time-definition2}
T^0(B_r) = \int_{-\infty}^{\infty}dt\, \chi_r[q+vt]= \frac{1}{v}\int_{-\infty}^{\infty}d\alpha\, \chi_r(\alpha)=\frac{2r}{v},
\end{equation} 
where for the second equality we have performed the change of variable $\alpha =q + vt$. 

In other terms, defining permanence as a time-integral over probabilities of presence, or as a difference between exit and entrance times, is in fact equivalent, when a notion of trajectory is available, and this remains clearly true also in more than one spatial dimension and for more complex dynamics than the free evolving one~\cite{Narnhofer, Sas1}.

\section{Quantum time of permanence}\label{Quantum time of permanence}

Let us come back to our concern, which is the definition of a proper time observable in QM. Thanks to  (\ref{sojourn-time-definition}), we can easily bypass the difficulty of the lack of a self-adjoint arrival time operator and define a quantum permanence time  as an integral over presence probabilities. Indeed, in QM the probability of presence of an entity (like an electron) inside a given region of space is a perfectly well defined quantity. More precisely, to the property ``The entity is inside the spatial ball $B_r$,'' we can associate an orthogonal projection operator $P_r$, such that if $|\varphi_t\rangle =e^{-\frac{i}{\hbar}H_0t}|\varphi\rangle$ is the state describing a free evolving quantum entity at time $t$, then
\begin{equation}
\label{probability of presence}
{\cal{P}}_{\varphi_t}(B_r)= \left\|P_r\varphi_t\right\|^2=\left\langle \varphi_t|P_r|\varphi_t\right\rangle =\int_{B_r} d^nx\, \left|\varphi_t(\textbf{x})\right|^2
\end{equation}
is the probability for the quantum entity to be found inside the ball $B_r$, at time $t$, following a measurement. Thus, using (\ref{sojourn-time-definition}), we can define the \textit{quantum permanence time} in $B_r$ by the integral:
\begin{equation}
\label{sojourn-time-definition-quantum}
T^0_\varphi(B_r)=\int_{-\infty}^{\infty}dt\, {\cal{P}}_{\varphi_t}(B_r).
\end{equation}

The conceptual validity of definition (\ref{sojourn-time-definition-quantum}) depends only on the conceptual validity of the probability (\ref{probability of presence}). And since the latter possesses a proper meaning in QM, the same must be true for the average (\ref{sojourn-time-definition-quantum}). In other terms, as a purely probabilistic statement, definition (\ref{sojourn-time-definition-quantum}) is independent of the details of the theory which underlies (\ref{probability of presence}), and in particular of the existence of a classical notion of trajectory. 

Let us describe some of the properties of the quantum time of permanence (\ref{sojourn-time-definition-quantum}), which can also be written as the average  
\begin{equation}
\label{free-sojourn-time-operator}
T^0_{\varphi}(B_r)=\left\langle \varphi |T^0(B_r)|\varphi\right\rangle
\end{equation}
of the operator
\begin{equation}
\label{free-sojourn-time-operator2}
T^0(B_r)=\int_{-\infty}^{\infty}dt\, e^{\frac{i}{\hbar}H_0t}P_r e^{-\frac{i}{\hbar}H_0t}
\end{equation}
over the state $|\varphi\rangle$ at time $t=0$. The operator (\ref{free-sojourn-time-operator2}) is known in the literature as the \textsl{free sojourn time operator}~\cite{Mar2}. We can observe that:
\begin{eqnarray}
\label{free-sojourn-time-operator-commutation with free evolution}
e^{\frac{i}{\hbar}H_0\alpha}T^0(B_r)&=&\int_{-\infty}^{\infty}dt\, e^{\frac{i}{\hbar}H_0(t+\alpha)}P_r e^{-\frac{i}{\hbar}H_0t} =\int_{-\infty}^{\infty}dt\, e^{\frac{i}{\hbar}H_0t}P_r e^{-\frac{i}{\hbar}H_0(t-\alpha)} \nonumber\\
&=& T^0(B_r)e^{\frac{i}{\hbar}H_0\alpha}.
\end{eqnarray}
Deriving (\ref{free-sojourn-time-operator-commutation with free evolution}) with respect to $\alpha$, then setting $\alpha =0$, we thus obtain:
\begin{equation}
\label{free-commutation-relation}
\left[H_0,T^0(B_r)\right]=0.
\end{equation}
In other terms, contrary to the (non self-adjoint) arrival time operator (\ref{time operator}), which formally obeys the canonical commutation relation with the free Hamiltonian, the  permanence (or sojourn) time operator $T^0(B_r)$ is fully compatible with the energy $H_0$ of the system and doesn't entertain with it an Heisenberg uncertainty relation. In particular, the previously mentioned Pauli's argument doesn't apply, and in fact one can show that $T^0(B_r)$ is a bona fide self-adjoint operator~\cite{Lavine, Jaw}. 

Let us also observe that the permanence time (\ref{free-sojourn-time-operator}) is finite only if the probability density $\left|\varphi_t(\textbf{x})\right|^2$ decreases sufficiently rapidly as $t\to\pm\infty$. If, for example, one choses for the state $\varphi$, at time $t=0$, a Gaussian wave packet, one can easily show that $\left|\varphi_t(\textbf{x})\right|^2=O\left(t^{-n}\right)$, so that $T^0_\varphi(B_r)$ is finite for $n\geq 2$, but infinite for $n=1$, and this for any choice of the initial velocity of the wave packet. 

This difference between the one-dimensional and higher dimensional cases can be explained in terms of the spreading of the wave packet, a purely quantum phenomenon with no analogues in CM: contrary to the case $n\geq 2$, in the $n=1$ situation the spreading of the wave packet increases at the same linear rate $t$ as the distance covered by the particle~\cite{Jaw}. 

In general, one can show that $T^0_\varphi(B_r)$ is a bounded operator for $n\geq 2$, and an unbounded, but densely defined operator, for $n=1$, typically on the set of states having no components near the zero of energy~\cite{Lavine, Jaw, Dambo}.

\section{An explicit calculation}\label{An explicit calculation}

To explicitly calculate the permanence time (\ref{free-sojourn-time-operator}), it is useful to introduce the simultaneous improper eigenvectors of $H_0$ and $\hat{\textbf{P}}= \textbf{P}/|\textbf{P}|$:
\begin{equation}
\label{impropereigenvectors}
H_0|E,\hat{\textbf{k}}\rangle=E |E,\hat{\textbf{k}}\rangle, \ \hat{\textbf{P}}|E,\hat{\textbf{k}}\rangle =\hat{\textbf{k}}|E,\hat{\textbf{k}}\rangle,
\end{equation}
where $E\in [0,\infty)$ and $\hat{\textbf{k}}\in {\cal S}^{n-1}$ (the unit sphere). They obey the relations of completeness
\begin{equation}
\label{completeness}
\int_0^\infty dE\, \int_{{\cal S}^{n-1}} d{\hat k}|E,\hat{\textbf{k}}\rangle\langle E,\hat{\textbf{k}}| =\mathbb{I}
\end{equation}
and orthogonality
\begin{equation}
\label{orthogonaity}
\langle E,\hat{\textbf{k}}| E^{\prime},\hat{\textbf{k}}^{\prime}\rangle = \delta(E-E^{\prime})\delta(\hat{\textbf{k}}-\hat{\textbf{k}}^{\prime})
\end{equation}
and their wave function is given by:
\begin{equation}
\label{planewaves}
\langle\textbf{x}|E,\hat{\textbf{k}}\rangle =\left(2\pi\hbar\right)^{-\frac{n}{2}}\sqrt{m}\left(2mE\right)^{\frac{n-2}{4}}e^{i\sqrt{2mE}\hat{\textbf{k}}\textbf{x}}.
\end{equation}
We denote by $\varphi(E)=\langle E|\varphi\rangle$ the vectors in $L^2({\cal S}^{n-1})$, at fixed energy, and by 
\begin{equation}
\label{fixed energy scalar product}
\langle\varphi(E)|\varphi^{\prime}(E)\rangle=\int_{{\cal S}^{n-1}}d\hat{k}\, \varphi^{\ast}(E,\hat{\textbf{k}})\varphi^{\prime}(E,\hat{\textbf{k}})
\end{equation}
the corresponding scalar product, where $\varphi(E,\hat{\textbf{k}})=\langle E,\hat{\textbf{k}}|\varphi\rangle$. 

For simplicity,  once more we limit ourselves to the one-dimensional case $n=1$. For a single spatial dimension, the unit sphere ${\cal S}^{0}$ is made only of two points, $\hat{\textbf{k}}=\pm 1$, and we can write $\hat{\textbf{P}}=P_+-P_-$, where
\begin{equation}
\label{one-dim proj-operators}
P_\pm=\int_0^\infty dE\, |E,\pm\rangle\langle E,\pm| \equiv |\pm\rangle\langle \pm|
\end{equation}
are the projection operators into the subspace of states of positive $(+)$ and negative $(-)$ momentum, respectively. 

For a free entity coming, say, from the left, i.e. $P_+ |\varphi\rangle =|\varphi\rangle$, we have:
\begin{eqnarray}
\label{free sojourn time calculation}
T^0_{\varphi}(B_r)&=&\left\langle \varphi |T^0(B_r)|\varphi\right\rangle =\!\!\int_0^\infty \!\!\!\! dE \int_0^\infty \!\!\!\! dE^{\prime}\, \varphi^{\ast}(E,+)\langle E,+|T^0(B_r)|E^{\prime},+\rangle\varphi(E^{\prime},+)\nonumber\\
&=&\!\!\int_0^\infty \!\!\!\! dE \int_0^\infty \!\!\!\! dE^{\prime}\, \varphi^{\ast}(E,+)\langle E,+|P_r|E^{\prime},+\rangle\varphi(E^{\prime},+) \int dt\, e^{\frac{i}{\hbar}(E-E^{\prime})t}\nonumber\\
&=&\int_0^\infty dE\, \langle +|T^0_E(B_r)|+\rangle |\varphi(E,+)|^2,
\end{eqnarray}
where for the last equality we have used the identity $\int dt\exp[\frac{i}{\hbar}(E-E^{\prime})t]=2\pi\hbar\delta(E-E^{\prime})$ and we have defined:
\begin{eqnarray}
\label{on-shell free sojourn time}
\langle +|T^0_E(B_r)|+\rangle &=& 2\pi\hbar \langle E,+|P_r|E,+\rangle = 2\pi\hbar\int_{-r}^r dx\, |\langle x|E,+\rangle|^2\nonumber\\
&=&2\pi\hbar\int_{-r}^r dx\,\left|\frac{1}{\sqrt{2\pi\hbar}}\sqrt{\frac{m}{\hbar k}}e^{ikx}\right|^2 = \frac{2r}{v},
\end{eqnarray}
with $v=\hbar k/m=\sqrt{2E/m}$. Finally, inserting (\ref{on-shell free sojourn time}) into (\ref{free sojourn time calculation}), and setting $\varphi(E)\equiv\varphi(E,+)$, we obtain that the one-dimensional quantum free permanence time in $[-r,r]$, for a particle coming from the left, is simply given by:
\begin{equation}
\label{one-dimensional free sojourn time}
T^0_{\varphi}(B_r)=\int_0^\infty dE\, \frac{2r}{v} |\varphi(E)|^2.
\end{equation}

In the limit of a wave packet sharply peaked about the energy $E$, i.e., in the limit 
\begin{equation}
\label{monoenergetic limit}
|\varphi(E^\prime)|^2\rightarrow \delta (E^\prime -E)
\end{equation} 
of a monoenergetic (but still square integrable) wave function, we obtain that the quantum one-dimensional incoming free sojourn time tends to
\begin{equation}
\label{one-dimensional free sojourn time fixed energy}
T^0_{\varphi}(B_r)\to\frac{2r}{v},
\end{equation}
i.e., to the classical expression (\ref{sojourn-time-definition2}); and of course, the same holds true for an entity moving from the right to the left.

\section{Interferences}\label{Interferences}

According to the above calculation, for a monoenergetic free entity coming from the left (or from the right), the quantum time of permanence coincides with the classical one (in the monoenergetic limit). This however cannot be true in general, because of the well known phenomenon of interference, which is typical of QM but totally absent in CM. Let us show how interferences manifest in the ambit of the one-dimensional free permanence time. 

For this, we recall that in QM interference terms arise as a consequence of the non compatibility of certain properties or, which is equivalent, of the non commutativity of certain observables. More precisely, consider two properties $a$ and $b$ and let $P_a$ and $P_b$ be the associated orthogonal projection operators. Let also $\bar{a}$ be the  inverse property of $a$, associated to the projector $P_{\bar a}=\mathbb{I}-P_a$. Then, we can write:
\begin{eqnarray}
\label{projectors1}
P_b &=& \left(P_a+P_{\bar a}\right)P_b\left(P_a+P_{\bar a}\right)\nonumber\\
&=&P_aP_bP_a + P_{\bar a}P_bP_{\bar a}+P_aP_bP_{\bar a}+P_{\bar a}P_bP_a.
\end{eqnarray}
Taking the expectation value of (\ref{projectors1}) over a state $|\varphi\rangle$, we find that the probability ${\cal P}_{\varphi}(b)=\langle\varphi|P_b|\varphi\rangle$ can be written as:
\begin{eqnarray}
\label{projectors2}
{\cal P}_{\varphi}(b) &=& {\cal P}_{\varphi}(a \;\textnormal{and then} \; b) + {\cal P}_{\varphi}(\bar{a}\; \textnormal{and then}\; b)\nonumber\\
&+& 2 \Re \langle \varphi |P_aP_bP_{\bar a}|\varphi\rangle,
\end{eqnarray}
where
\begin{equation}
\label{jointprobability1}	
{\cal P}_{\varphi}(a \;\textnormal{and then} \; b)=\langle \varphi |P_aP_bP_a|\varphi\rangle
\end{equation}
is the expectation value (which lies between $0$ and $1$) of the self-adjoint operator $P_aP_bP_a$, which can roughly be interpreted as corresponding to a measure of property $a$ immediately followed by a measure of property $b$. Similarly, 
\begin{equation}
\label{jointprobability2}	
{\cal P}_{\varphi}(\bar{a} \;\textnormal{and then} \; b)=\langle \varphi |P_{\bar{a}}P_bP_{\bar{a}}|\varphi\rangle	
\end{equation}
is the expectation value of the self-adjoint operator $P_{\bar{a}}P_bP_{\bar{a}}$, corresponding to a measure of property $\bar{a}$ immediately followed by a measure of property $b$. When properties $a$ and $b$ are compatible (i.e., the associated orthogonal projection operators commute), the last term in (\ref{projectors2}) is zero and one finds that:
\begin{equation}
\label{total probability}
{\cal P}_{\varphi}(b)={\cal P}_{\varphi}(a \;\textnormal{and then} \; b) + {\cal P}_{\varphi}(\bar{a} \;\textnormal{and then} \; b),
\end{equation}
which is the theorem of total probability of classical probability theory. 

In this case ${\cal P}_{\varphi}(a \;\textnormal{and then} \; b)$ and  ${\cal P}_{\varphi}(\bar{a} \;\textnormal{and then} \; b)$ can be interpreted as the \textit{joint} probabilities associated to the \textit{meet} properties $ab$ and $\bar{a}b$, respectively. However, if $a$ and $b$ are not compatible, the last term in (\ref{projectors2}), which is an interference term, will in general be different from zero, and one cannot anymore interpret ${\cal P}_{\varphi}(a \;\textnormal{and then} \; b)$ and ${\cal P}_{\varphi}(\bar{a} \;\textnormal{and then} \; b)$ as joint probabilities, at least not in the usual sense of classical probability theory.

To show how interferences manifest at the level of the quantum permanence time, we let $b$ be the property ``The entity is inside the ball $B_r$'', associated to the projector $P_r$, $a$ the property ``The entity has positive momentum'', associated to the projector $P_+$, and $\bar{a}$ the property  ``The entity has negative momentum'', associated to the projector $P_-$. We let also the state describing the quantum entity at time $t=0$, be described by a superposition $|\chi\rangle=(|\varphi_1\rangle+|\varphi_2\rangle)/\sqrt{2}$, where $|\varphi_1\rangle$ is a normalized state with only positive momentum, i.e., $P_+|\varphi_1\rangle=|\varphi_1\rangle$, and $|\varphi_2\rangle$ a normalized state with only negative momentum, i.e., $P_-|\varphi_2\rangle=|\varphi_2\rangle$. (Being $|\varphi_1\rangle$ and $|\varphi_2\rangle$ orthognal, $|\chi\rangle$ is duly normalized to $1$). Then, (\ref{projectors2}) becomes:
\begin{equation}
\label{interference1}
{\cal P}_{\chi}(B_r) = \frac{1}{2}\left[{\cal P}_{\varphi_1}(B_r) + {\cal P}_{\varphi_2}(B_r)\right]+\Re \int_{-r}^r dx\, \varphi_1^\ast(x)\varphi_2(x),
\end{equation}
which makes even more evident the interpretation of the last term in (\ref{interference1}) as an interference term. Therefore, considering a free evolving state $|\chi_t\rangle =e^{-\frac{i}{\hbar}H_0t}|\chi\rangle$, we find for the permanence time:
\begin{equation}
\label{interference2}
T^0_{\chi}(B_r) = \frac{1}{2}\left[T^0_{\varphi_1}(B_r) + T^0_{\varphi_2}(B_r)\right]+  \Re \langle\varphi_1 |T^0(B_r)|\varphi_2\rangle.
\end{equation}

More explicitly, if for instance we choose for the initial state $|\chi\rangle$ an odd function of the momentum\footnote{If we consider the one-dimensional Schr\"odinger equation as a radial equation ($x\geq 0$), then $\chi$ describes a spherical $3$-dimensional wave of zero angular momentum ($s$-wave).}, i.e., $\varphi_1(E,+)=-\varphi_2(E,-)\equiv \varphi(E)$, then the last interference term of (\ref{interference2}) is given by the following oscillating contribution: 
\begin{eqnarray}
\label{interference3}
\Re \langle\varphi_1 |T^0(B_r)\varphi_2\rangle &=& \Re \int_0^\infty \!\!\!\! dE\, \varphi_1^\ast(E,+)\langle +|T^0_E(B_r)|-\rangle \varphi_2(E,-)\nonumber\\
&=& -\Re \int_0^\infty dE\, |\varphi(E)|^2\int_{-r}^r dx\, \frac{m}{\hbar k}e^{-2ikx}\nonumber\\
&=&-\int_0^\infty dE\, \frac{\hbar}{2E}\sin(2kr)|\varphi(E)|^2.
\end{eqnarray}
Thus, the free permanence time becomes:
\begin{equation}
\label{free-sojourn-time-s-wave}
T^0_{\chi}(B_r)=\int_0^\infty dE\,\left[\frac{2r}{v}-\frac{\hbar}{2E}\sin(2kr)\right]|\varphi(E)|^2.
\end{equation}
Finally, taking for $|\varphi(E)|^2$ the monoenergetic limit (\ref{monoenergetic limit}), one finds that the permanence time of a free quantum particle whose initial state is described by an odd function of momentum, tends to
\begin{equation}
\label{free-sojourn-time-s-wave monoenergetic}
T^0_{\chi}(B_r)\to \frac{2r}{v}-\frac{\hbar}{2E}\sin(2kr),
\end{equation}
i.e., to the classical ``free-flight'' value plus an interference (oscillating with $r$) contribution, with no classical analogue.

\section{Non-spatiality}\label{Non-spatiality}

The above result should not surprise us: interferences are ubiquitous in QM, being the direct consequence of the superposition principle. And of course, there is no reason to believe that time of permanence observables would be free from that typical quantum phenomenon. But, is this the end of the story? Do we have to conclude that the classical concept of time of permanence (or time of sojourn) straightforwardly generalizes to the quantum realm, and that although a classical notion of trajectory is not readily disposable in QM, a classical notion of time of permanence nevertheless is? 

As we observed, times of permanence can be defined as average quantities, i.e., sums over probabilities of presence. And, if we understand these sums as pure probabilistic statements, they clearly remain conceptually consistent also when a classical notion of trajectory is not available. But, in what sense a concept of trajectory is not available in QM, and what are the real consequences of this for a correct interpretation of (\ref{free-sojourn-time-operator}) as a time of permanence?

To answer this fundamental question, we start by observing that the very notion of permanence time, if taken seriously, already implies that the classical concept of trajectory has to be fully abandoned in QM. For this, we consider a time-interval $[t_1,t_2]$, and define by the integral 
\begin{equation}
\label{free sojourn-time finite time}
T^0_\varphi(B_r;[t_1,t_2]) = \int_{t_1}^{t_2}dt\, {\cal P}_{\varphi_t}(B_r) =\langle \varphi |T(B_r;[t_1,t_2])|\varphi\rangle \leq t_2 - t_1,
\end{equation} 
the average time spent by the quantum entity inside $B_r$, during the time-interval $[t_1,t_2]$, where 
\begin{equation}
\label{sojourn-time operator finite time}
T^0(B_r;[t_1,t_2]) = \int_{t_1}^{t_2}dt\, e^{\frac{i}{\hbar}H_0t}P_r e^{-\frac{i}{\hbar}H_0t}
\end{equation}
is the (free) permanence time operator (\ref{free-sojourn-time-operator2}), restricted to the time-interval $[t_1,t_2]$. As it is a bona fide self-adjoint operator, according to the spectral theorem we know that there exist a projection-valued measure $F^0(B_r;[t_1,t_2]; \cdot)$, such that (\ref{sojourn-time operator finite time}) can be rewritten in the diagonal form 
\begin{equation}
\label{sojourn-time operator finite time, spectral resolution}
T^0(B_r;[t_1,t_2]) = \int_{\mathbb{R}} F^0(B_r;[t_1,t_2]; d\tau)\, \tau.
\end{equation}

Now, although $T^0(B_r;[t_1,t_2])$ is  self-adjoint, and therefore is a perfectly well-defined quantum observable, it is certainly not an observable in the conventional sense, as it doesn't correspond to an instantaneous measurement, but, rather, to a continuous measurement in the limit of zero-disturbance\footnote{Such a continuous measurement can be implemented by specific clocks, like the spin-clock, consisting in locally applying a constant magnetic field in the $B_r$-region, to activate and deactivate the spin precession at the entry and exit of it. Then, in the limit of a zero-field strength (i.e., in the limit of a zero-perturbation of the spin-clock on the quantum entity's movement), it can be proved that the total accumulated precession angle coincides with the permanence time (\ref{sojourn-time-definition-quantum}) of the quantum entity inside $B_r$~\cite{JawWard3,Mar3,Max}.}. However, one can reasonably extend the usual Born rule also to $T^0(B_r;[t_1,t_2])$, taking seriously its interpretation as a time of permanence observable, hence interpreting the associated projection-valued measure in the usual probabilistic sense. 

More precisely, given a Borel subset $\Delta \subseteq \mathbb{R}$, $F^0(B_r,[t_1,t_2];\Delta)$ is to be interpreted as the projection operator into the set of states that, in the course of their (free) evolution, spend inside $B_r$, during the time interval $[t_1,t_2]$, amounts of time whose values are in $\Delta$. In other terms, the average
\begin{equation}
\label{probability projection valued measure}
{\cal P}_{\varphi}(B_r;[t_1,t_2];\Delta) =\langle \varphi |F^0(B_r;[t_1,t_2];\Delta)|\varphi\rangle
\end{equation}
is the probability that a free evolving quantum entity with initial state $|\varphi\rangle$, sojourns in $B_r$, during the time-interval $[t_1,t_2]$, an amount of time belonging to the set $\Delta$. 

Then, setting $\Delta =\{0\}$, ${\cal P}_{\varphi}(B_r;[t_1,t_2];\{0\})$ is the probability for the free entity of spending a \textit{zero} amount of time in $B_r$, during the time-interval $[t_1,t_2]$. Said it differently, it corresponds to the probability for the entity to not enter, for any measurable amount of time, the spatial region $B_r$, during the time-interval $[t_1,t_2]$. 

Now, the puzzling result that was proven by Jaworski~\cite{Jaw}, is that for any choice of $|\varphi\rangle$ and time-interval $[t_1,t_2]$, such a probability is always equal to zero. In other terms, there are no eigenstates of the permanence time operator (\ref{sojourn-time operator finite time}) corresponding to the zero eigenvalue. To put it differently, this means that the quantum entity will always spend (with probability $1$) some time in $B_r$, during whatever time-interval $[t_1,t_2]$, and this independently of the choice of its initial condition.

So, if we take seriously the interpretation of (\ref{free sojourn-time finite time}) as a measure of the time spent inside $B_r$, during the time-interval $[t_1,t_2]$, and if we assume that the quantum entity is a \textit{spatial} entity, that is, an entity existing and evolving inside our three-dimensional physical space, we are faced with an apparent paradox. Indeed, if the entity is a particle, i.e., a \textit{local} corpuscle, then taking a ball $B_r$ of arbitrary small radius $r$, a time-interval $[t_1,t_2]$ with $t_2$ arbitrary close to $t_1$, and a state that has been prepared in such a way that, at time $t_1$, it is localized at an arbitrary astronomical distance far away from the origin, we clearly expect that, however strange, erratic and speedy would be the free quantum ``particle'' displacements in space, under such conditions the time it spends in $B_r$, during the (infinitesimal) time-interval $[t_1,t_2]$, has to be equal to zero. But, as we said, this expectation is false, and therefore the hypothesis that the quantum entity is a \textit{local} entity is not tenable and must be abandoned. 

Then, let us assume that, on the contrary, it is a \textit{non-local} entity, i.e., an entity that, somehow, is spread all over space. In this case it becomes relatively easy to understand that zero cannot be an eigenvalue of the sojourn time operator, as the quantum entity would possess the remarkable property of being present, in every moment, in every region of space. However, setting $t_1=-\infty$ and $t_2=\infty$, we would then expect in this case the permanence time (\ref{free sojourn-time finite time}) to be infinite. But this is again false, as  we know that (\ref{free-sojourn-time-operator2}) is a bounded operator.

We are thus forced to conclude that the crucial point is not the locality or non-locality of the quantum particle, but its presumed \textit{spatiality}. The only possible conclusion is that if a microscopic entity can manifest as a non-local entity, it is because it is first of all a \textit{non-spatial} entity, i.e., an entity that sojourns most of its time in a space that is not our three-dimensional Euclidean space~\cite{Aer, Sas3}. And for that reason, a (microscopic) quantum entity shouldn't be called ``particle'', as to be such it should possess at least the attribute of spatiality\footnote{In fact, a microscopic (quantum) entity doesn't possess many other fundamental attributes that are usually associated to a particle, like for instance the one of \textit{individuality}; see Ref.~\onlinecite{Sas3} and the references cited therein.}. 

If a quantum entity doesn't possess, in general, a position in space, as ``having a spatial position'' is just a property (most of the time ephemeral) that is created during a measurement process, it is clearly improper to refer to (\ref{free sojourn-time finite time}) as a permanence time, as the term ``permanence'' refers to the property of remaining (or sojourning) in the \textit{spatial} region $B_r$, whereas the quantum entity is a \textit{non-spatial} entity, that is, an entity that doesn't sojourn in physical space! 

But then, if we nevertheless consider that the self-adjoint observable (\ref{free-sojourn-time-operator2}) is telling us something about the reality of the quantum world, what is it exactly?  In other terms, how should we interpret the sum (\ref{sojourn-time-definition-quantum})? Before answering these questions, let us briefly explore in the next section a particular approach to reality, called the \textit{creation-discovery view}.

\section{The creation-discovery view}\label{The creation-discovery view}

In the previous section we used the spectral properties of the permanence time operator to show that microscopic quantum entities, like for instance electrons, are not permanently present in our Euclidean three-dimensional space. In a recent paper in this journal~\cite{ Sas3}, we reached the same conclusion by combining \textit{Heisenberg's uncertainty principle} with the notion of \textit{element of reality}, as firstly introduced by Einstein Podolsky and Rosen and further developed in the ambit of the so-called Geneva-Brussels approach to the foundations of quantum mechanics (see Ref.~\onlinecite{ Sas3} and the references cited therein). 

Let us remind however  that strong emphasis on the non-spatiality of quantum entities was already given in the last decades by D. Aerts and his collaborators in the Brussels' group. Quoting Aerts (Ref.~\onlinecite{Aer}, page 178): ``Reality is not contained within space. Space is a momentaneous crystallization of a theatre for reality where the motions and interactions of the macroscopic material and energetic entities take place. But other entities - like quantum entities for example - `take place' outside space, or - and this would be another way of saying the same thing - within a space that is not the three dimensional Euclidean space.''

The non-spatiality of quantum entities can also be understood as a consequence of the application to the interpretation of QM of a very general conceptual framework, called the \textit{creation-discovery view} (CDV), which was developed in the past years by D. Aerts and his collaborators in Brussels. Let us briefly recall in this section some of the important ideas of the CDV, keeping in mind that this conceptual language is much richer and subtler than what can be appreciated by our brief mention here, which has the only purpose of motivating and clarifying the introduction, in the next section, of the notion of \textit{total availability}, which follows intrinsically from it. We therefore refer the interested reader to Refs. \onlinecite{Aer, Aer2, Aer3}, and the references cited therein, for a more complete exposition.

The CDV has been developed to successfully integrate in a coherent conceptual framework our experimental and theoretical knowledge about classical and quantum systems, thus providing a general and articulated form of realism, which not only acknowledges that there are things ``out there,'' existing regardless of our acts of observation, but also that our observations are not always without consequences, insomuch that they can go as far as to literally create the very properties which are being observed. 

In our active role of \textit{participators} of reality, we are constantly interacting with the countless entities that are populating it, and these interactions form the basis of our \textit{experiences}, of which the measurements we perform in our physical laboratories are just special cases (the participator being then constituted by the scientist together with his apparatus and the experience being usually called an experiment). 

Now, in the same way as in our ordinary language we distinguish between \textit{verbs} and \textit{substantives}, in our experiences (or experiments) we can distinguish between \textit{creation} and \textit{discovery} aspects. The discovery aspects describe those properties in reality that are actual before the experience, whereas the creation aspects correspond to the new properties that are created during its execution, as a consequence of the interaction of the participator with the elements of reality (not necessarily spatial) that take part in the experience. 

To fix ideas, let us consider the simple experience of drinking a cup of tea in a tearoom. We can observe that many entities populate the setting of such an experience. There is of course the cup of tea, filled with some tasty tea, but also the table, on which the cup is placed, the chairs, the air filling the volume of the room, the other people around, with their thoughts and emotions, and so on. All these elements exist independently from the participator. Aerts call them \textit{happenings}, to distinguish them from the usual \textit{events} of relativity theory and to emphasize the fact that not every happening necessarily happens in space (and/or time), as it is the case for microscopic quantum entities or, in our example, for the thoughts and emotions of the persons present in the tearoom, which belong to more abstract mental and emotional spaces.

From the viewpoint of the participator, all these happenings happen at once in his reality. And for the very fact that they happen, each of them can possibly be selected to participate to one of his experiences. By definition, an experience lived by the participator is the combination of one of his acts of creation (usually described by a verb) with one of its \textit{available} happenings (usually described by a substantive). 

In our example, the creation aspect is simply the participator's action of taking the cup of tea in his hands and drinking its contents; an action which is fully under his control. On the other hand, the discovery aspect of the experience is that specific happening of his reality -- called the full cup of tea -- which lends itself to his creation, through his action of taking it and drinking a certain amount of its liquid content. The outcome of the experience is the creation of a new entity, that wasn't actually existing (i.e., happening) prior to its execution, which can either be a ``totally empty cup of tea,'' or a ``partially empty cup of tea,'' according to the amount of liquid the participator is able to swallow in one go.

It is worth observing that whereas an arbitrarily large number of happenings can all happen at once, only one at a time can participate to a participator's experience, as experiences can only be experienced at once, in the participator's present moment. Nevertheless, all the available happenings do contribute to the construction of the participator's reality, and this is so because they could have been the happening aspect of his present experience, if he had decided to act differently in his past.

This means that what we call reality (and more precisely our personal present reality) is a construction about what is possible, i.e., about the experiences we could have lived in replacement of our present one, if only we would have decided so in our past. And this also means that our construction of reality depends essentially on two factors: (1) our \textit{personal power} in performing certain act of creations and (2) the \textit{availability} of the happenings that can possibly participate in the experiences we have the power to perform. 

Another important aspect in the creation-discovery view that is worth mentioning is the one of the predictability or unpredictability of the outcome of an experience, or experiment. For instance, let us assume that before drinking the cup of tea the participator has already decided to empty the cup in one go. Then, we can say that the outcome of the drinking experiment is fully under his control and will result with certainty in the creation of the entity called the ``totally empty cup of tea.'' Alternatively, the participator could have decided to only bring the cup to his nose and smell the tea, in order to gather knowledge about its aroma and temperature. In this case, the full cup entity would remain unaffected by his action, which corresponds to a pure observational process where only knowledge about the state of the entity is gathered during its course; a situation which is typical of classical physics experiments. 

But predictability is not necessarily the rule when the participator (and his measuring apparatus) is not fully controlling all the variables intervening in the interaction, so that the outcome will not in general be predetermined. For instance, although the participator may know everything about the state of the full cup of tea (its size, temperature, the volume of the liquid content, etc), if he hasn't decided in advance how  the drinking interaction will have exactly to be conducted, then a certain number of contextual factors will be allowed to influence his drinking process, possibly resulting in a ``totally empty cup of tea,'' or a ``partially empty cup of tea,'' or even something else (like a broken cup of tea). 

According to the creation-discovery view, quantum systems are such because in certain experiments (called measurements) the participator cannot control all those fluctuations in the interaction between the system and the measuring apparatus, that can effect the outcome. This picture has been clearly substantiated in the so-called \textit{hidden measurement approach}~\cite{Aer, Aer2, Aer3}, where it has been shown that the typical behavior of quantum systems can be reproduced by quite conventional macroscopic entities (called \textit{quantum machines}), whenever a selection mechanism is at work during the experiment, picking out a specific measurement among different (deterministic) ``hidden measurements,'' therefore selecting a specific outcome. 

If such a selection mechanism is not under the control of the experimenter, this will result in the typical quantum probabilities which, contrary to Kolomogorovian classical probabilities, cannot to be associated to a lack of knowledge about what is already actually existing prior to the measurement (like the state and properties of the system under consideration), but to a lack of knowledge about what is exactly going on during the interaction between the system and the measuring apparatus.  

In other terms, the distinction between classical and quantum probabilities would be just a distinction between discovering what is already there and creating what is still not there, by  means of an experiment (i.e., a measurement process) whose internal dynamics is not controlled (or fully controlled) by the experimenter.

Obviously, in the case of our simple example, every participator would agree that only the full cup of tea actually exists prior to the ``drinking interaction'' and that the totally empty cup or the partially empty cup are only potentially existing. Therefore, to no one would come the idea to measure (i.e., to test) the total emptiness or partial emptiness of the full cup of tea, because the totally empty and partially empty cups are only potentially present, and not actually present in space. In other terms, they are not happenings belonging to the present reality of the participator, but only possible future happenings of its future reality, that can be created during his drinking process.  

But then, taking seriously the conceptual framework of the creation-discovery view, we may realize, as emphasized by Aerts~\cite{Aer, Aer2, Aer3}, that when we pretend measuring the position of, say, an electron, we are probably committing the same conceptual mistake as if we would pretend measuring the total emptiness of a full cup of tea. Indeed, measuring the position of an electron is about measuring the \textit{spatial} position of a \textit{non-spatial} entity, which is obviously a contradiction in terms.

\section{Total availability}\label{Total availability}

In the previous section we have recalled some of the central ideas of the creation-discovery view, according to which our reality consists of those entities (happenings) that are \textit{available} to us, in the sense of being available to our \textit{experiences}, which are essentially \textit{creation-discovery} processes, and that what we usually call classical observations are experiences of pure discovery (i.e., of discovery of what is already manifest), whereas quantum ``observations'' are experiences of creation of what isn't yet manifest, but could be manifested by means of the ``measuring'' process.

In the case of our concern, the entity in question is a microscopic quantum entity and the creation aspect of the experience is the one of manifesting a spatial localization, through the interacting with a local macroscopic measuring apparatus. As we explained in the previous two sections, contrary to the case of a classical macroscopic object, the spatial localization of a quantum entity doesn't exist prior to the observational process (or it exists, but only in a potential sense). Consequently, when measuring the spatial localization of the non-spatial quantum entity (evidently an inappropriate expression), we are in fact trying to create such a localization, and our attempt may or may not be successful, as we don't control all the variables of the experimental process. And the \textit{relative frequency of success} with which we can do this, is obviously a measure of the (degree of) \textit{availability} of the quantum entity in participating in such a spatial experience and produce a successful result.

In other terms, what we are here proposing is to distinguish availability in a general, unqualified sense (as described in the previous section), from availability in a more specific, qualified sense, in relation to the outcome of a certain experiment. Consequently, we are also suggesting to understand the quantum probability ${\cal P}_{\varphi_t}(B_r)$ as a measure of the (degree of) \textit{availability}, at time $t$, of the non-spatial quantum entity described by $|\varphi_t\rangle$, in lending itself to an interaction with a measuring apparatus, in order to manifest (i.e., to create) a (temporary) spatial localization inside the ball $B_r$.

Let us observe that because of its very existence, a quantum entity is always available, in a general sense, to participate to whatever experience, irrespective of its outcome. In other terms, its (degree of) availability to participate to whatever experience is equal to $1$, and will remain such in the course of its evolution. Mathematically, this is expressed by the fact that the initial condition of the quantum entity is described by a vector of the Hilbert space that is normalized to one, and that the evolution is unitary, so that $\langle\varphi_t|\varphi_t\rangle =1$, for all $t\in\mathbb{R}$. 

Of course, one can imagine situations where even this general availability wouldn't remain constant in the course of time. Indeed, we can consider that in the same way as existing entities have once been created, and therefore have become part of our reality, they can as well be destroyed, and thus cease to be an available part of our possible experiences. In the conventional non-relativistic quantum formalism, the latter process can for instance be modeled by the introduction of a purely imaginary (dissipative) term in the Hamiltonian, which in the present free case would then become: $H = H_0 +i\lambda W$, with $\lambda$ a real coupling constant and $W$ a given self-adjoint operator. Then, the evolution is not anymore unitary, but given by a semi-group of contractions~\cite{Mar1}: 
\begin{equation}
\label{contraction}
U(t,0)=
\begin{cases}
e^{-\frac{i}{\hbar}Ht}, & t>0 \\
e^{-\frac{i}{\hbar}H^\ast t}, & t<0. 
\end{cases}
\end{equation}
The scalar product $\langle\varphi_t |\varphi_t\rangle$ can then be interpreted as the probability for the particle to still exist, i.e., to still be avaialble to participate, at time $t$, to whatever experience, with whatever outcome. For instance, in the simple homogeneous case $W=\mathbb{I}$, (\ref{contraction}) can easily be integrated to give the exponential law: 
\begin{equation}
\label{exponential-law}
\langle\varphi_t |\varphi_t\rangle = e^{-2 \frac{\lambda}{\hbar}|t|},
\end{equation}
showing that the total availability of the quantum entity exponentially decreases as time passes by.

Having said that, let us limit our considerations, for simplicity, to the ideal case of quantum entities that we assume have an infinite life span, i.e., such that their unqualified (degree of) availability remains constantly equal to $1$, in the course of their evolution. Then, according to our previous considerations, the proper interpretation of the quantum permanence ``time''  
\begin{equation}
\label{total spatial availability}
T^0_\varphi(B_r) = \int_{-\infty}^\infty dt\, {\cal P}_{\varphi_t}(B_r)
\end{equation}
is as follows: it is not the time spent, on average, by the free-evolving entity inside $B_r$, but its \textit{total availability} in $B_r$, that is, its \textit{total availability in lending itself to the creation of a spatial localization inside $B_r$, by means of an interaction with a measuring apparatus}.
  
This means that the classical concept of \textit{time of permanence} has to be replaced in QM by the more general concept of \textit{total availability}, that is, the total availability of a quantum entity in being part of an experience the outcome of which is the creation of a temporary localization in a given region of space (which, in the present case, is $B_r$).

\section{Total availability shift}\label{Availability change}

Considering the limit $r\to\infty$, i.e., $B_r\to\mathbb{R}^n$, then $P_r\to\mathbb{I}$ and ${\cal P}_{\varphi_t}(B_r)\to 1$, for all $t$, so that $T^0_\varphi(B_r)\to\infty$. This means that the \textit{total availability} of a quantum entity in being localized in the whole three-dimensional space  is infinite. Of course, this is so because we are dealing here with hypothetical eternal entities. If the quantum entity would have a finite existence, then its total spatial availability would be finite. For instance, considering the exponentially decaying example (\ref{exponential-law}), we would have in this case that
\begin{equation}
\label{life span}
\lim_{r\to\infty}T^0_\varphi(B_r) = \int_{-\infty}^\infty dt\, e^{-2 \frac{\lambda}{\hbar}|t|}=\frac{\hbar}{\lambda}<\infty,
\end{equation}
which, in classical terms, one would interpret as the total life span of the entity, i.e., its average time of permanence in our physical space. 

The infinite total spatial availability of a quantum entity evolving according to a unitary evolution corresponds, in classical terms, to the infinite time of permanence of an eternal classical particle in three-dimensional space. 

Clearly, although so far we have only considered, for sake of simplicity, free evolving entities, we could have extended our discussion to entities evolving in the presence of a force-field, whose evolution is for instance governed by the Hamiltonian $H=H_0 +V$, where $V$ is the potential energy. Then, we would have found that also for an interacting entity the total spatial availability is infinite. It is therefore natural to ask if, in some way, it would be possible to extract some useful information from these infinities. 

This can certainly be done, for instance by conveniently comparing them. A possibility is to subtract from the total spatial availability of the interacting entity, the availability of an identical but free evolving entity, having same initial condition in the remote past (or, equivalently, in the distant future). 

Thinking again in classical terms, this brings us to the well-known concept of time-delay. Let us recall that in physics one usually associates a time-delay to a \textit{scattering} particle moving in the presence of a force-field. Time-delay then measures the excess or defect of time the particle spends in the interaction region, when its movement is compared to that of a free particle, subject to similar initial conditions in the remote past (or final conditions in the distant future). The knowledge of time-delay provides information about the effects of the interaction: generally speaking, a positive time-delay corresponds to an effect of deceleration; a positive large time-delay corresponds to the formation of a metastable, quasi-bound state; an infinite positive time-delay corresponds to the capture of the particle by the interaction; finally, a negative time-delay indicates that the particle has been accelerated by the effects of the interaction.

In classical physics, one can define the notion of time-delay as a difference of arrival times or, equivalently, as a difference of permanence times~\cite{Sas1, Max}. In quantum physics, on the contrary, only permanence times can be used in the time-delay definition. However, as we explained, quantum permanence times are not ``times'': they are quantifiers of the total availability of quantum entities in being localized in given regions of space. Therefore, the correct interpretation of the quantum generalization of the classical notion of time-delay should be the following: it is not a measure of the excess or defect time globally spent by the quantum entity in the interaction region, but a measure of the \textit{total (spatial) availability shift} it experiences as a consequence of the interaction.

Time-delay has been an extensively investigated subject, since the early days of quantum physics~\cite{Mar2, Max}, and analyzing the physical and mathematical content of this important notion would go far beyond the scope of the present note. Our intention here was only to propose a new conceptual interpretation for this classical quantity, in accordance with the non-spatial nature of quantum entities. 

Nevertheless, let us briefly describe, for completeness, how time-delay is usually defined in QM, but let us do so using our new conceptual terminology. For this, let $|\psi_t\rangle$ be the scattering state describing a quantum entity whose evolution is governed by the total Hamiltonian $H=H_0+V$. By definition, it has asymptotic behavior
\begin{equation}
\label{asymptotic-quantum}
|\psi_t\rangle=
\begin{cases}
e^{-\frac{i}{\hbar}H_0t}|\varphi\rangle, & \text{$t\rightarrow -\infty$} \\
e^{-\frac{i}{\hbar}H_0t}S|\varphi\rangle, & \text{$t\rightarrow +\infty$,} 
\end{cases}
\end{equation}
where $|\varphi\rangle$ is the so-called incoming state at time $t=0$, and $S$ the unitary scattering operator, that can be expressed in terms of the isometric wave operators $\Omega_\pm$ by the relation  $S=\Omega_+^\dagger\Omega_-$. The scattering state at time $t$ can be written as $|\psi_t\rangle = e^{-\frac{i}{\hbar}Ht}\Omega_-|\varphi\rangle = \Omega_-e^{-\frac{i}{\hbar}H_0t}|\varphi\rangle$, where we have used the intertwining property $H\Omega_\pm = \Omega_\pm H_0$, from which it also follows that the scattering operator is compatible with the free evolution, i.e., $H_0S=SH_0$. 

The total availability of the quantum interacting entity in $B_r$ is then given by 
\begin{equation}
\label{sojourn-time-operator}
T_\varphi(B_r)=\int_{-\infty}^{\infty}dt\, {\cal{P}}_{\psi_t}(B_r)=\left\langle \varphi |T(B_r)|\varphi\right\rangle,
\end{equation}
where $T(B_r)$ is the (self-adjoint) \textit{total availability operator}
\begin{equation}
\label{sojourn-time-operator2}
T(B_r)=\int_{-\infty}^{\infty}dt\, e^{\frac{i}{\hbar}H_0t}\Omega_-^\dagger P_r\Omega_- e^{-\frac{i}{\hbar}H_0t}.
\end{equation}
Repeating the same argument as for the free case, we find that it also commutes with the free evolution, i.e.,
\begin{equation}
\label{commutation-relation}
\left[H_0,T(B_r)\right]=0,
\end{equation}
and thus possesses on the energy shell matrix elements. 

The \textit{total availability shift} $\Delta T_\varphi$ of the quantum entity, induced by the ``switching on'' of the interaction (the equivalent of the classical notion of time-delay) is then given by the limit of the difference
\begin{equation}
\label{change in availability}
\Delta T_\varphi = \lim_{r\to\infty}\left[T_\varphi(B_r) -T_\varphi^0(B_r)\right].
\end{equation}

To illustrate the content of the above limit, let us consider once more the simple case of a one-dimensional scattering entity with an incoming state of positive momentum. As for (\ref{free sojourn time calculation}), an explicit calculation gives~\cite{Max}: 
\begin{equation}
\label{sojourn time calculation}
T_\varphi(B_r)=\int_0^\infty dE\, \langle +|T_E(B_r)|+\rangle |\varphi(E)|^2,
\end{equation}
where
\begin{align}
\label{sojourn time calculation2}
\langle +| &T_E(B_r)|+\rangle = |T_E|^2\hbar\frac{d\alpha^T_E}{dE} + |L_E|^2\hbar\frac{d\alpha^L_E}{dE}\nonumber\\
&+ \frac{2r}{v}+\frac{\hbar}{2E}|L_E|\sin\left(\alpha^L_E +2kr\right) + o(1),
\end{align}
as $r\to\infty$, with $T_{E}=|T_{E}|e^{i\alpha^{T}_{E}}$ and $L_{E}=|L_{E}|e^{i\alpha^{L}_{E}}$ the transmission and reflection amplitudes, respectively. We can observe that the last term in (\ref{sojourn time calculation2}) is again of an interference nature, but this time is due to the superposition of incoming and reflected waves. 

Setting $T_E = 1$ in (\ref{sojourn time calculation2}), we clearly recover the free expression (\ref{on-shell free sojourn time}). So, in the  difference (\ref{change in availability}), the divergent terms, growing linearly with $r$, duly cancel. On the other hand, the oscillating interference term in (\ref{sojourn time calculation2}) doesn't contribute to the $r\to\infty$ limit, because of the Riemann-Lebesgue Lemma. Therefore, one finds that the total availability shift of the quantum entity is given by
\begin{equation}
\label{quantum local time-delay one dimension}
\Delta T_\varphi =\int_0^\infty dE\, \langle +|\Delta T_E|+\rangle |\varphi(E)|^2,
\end{equation}
where
\begin{equation}
\label{quantum local time-delay one dimension2}
\langle +|\Delta T_E|+\rangle = |T_E|^2\hbar\frac{d\alpha^T_E}{dE} + |L_E|^2\hbar\frac{d\alpha^L_E}{dE}.
\end{equation}

Although we have here considered a simple one-dimensional example, it is possible to derive the above limit in more general contexts, using a number of different mathematical methods~\cite{Mar2, Max}. In general, one can show that for sufficiently regular interactions and incoming states, the limit (\ref{change in availability}) exists and is equal to~\cite{Mar2}
\begin{equation}
\label{time-delay general}
\Delta T_\varphi =\int_0^\infty dE\, \langle\varphi(E)|\Delta T_E|\varphi(E)\rangle,
\end{equation}
where the \textit{total availability shift operator} $\Delta T_E$ on the energy shell is given by
\begin{equation}
\label{E-W}
\Delta T_E = -i\hbar S^\dagger_E\frac{dS_E}{dE}
\end{equation}
and is usually referred to as the \textit{Eisenbud-Wigner operator} in the literature~\cite{Mar2, Max}.

\section{Discussion}\label{discussion}

If it is true, as we emphasized in the beginning of this article, that our idea of time is dependent upon our classical observation of macroscopic entities moving along trajectories in the three-dimensional space, and if it is also true, as hypothesized by Aerts, that (Ref.~\onlinecite{Aer}, page 135): ``[$\cdots$] quantum entities are not permanently present in space, and that, when a quantum entity is detected in such a non-spatial state, it is `dragged' or `sucked up' into space by the detection system,'' then we are forced to recognize that temporal notions like \textit{time of permanence} and \textit{delay time} are classical concepts that need to be upgraded in order to remain fully consistent, also in relation to quantum entities.  

In the present work we have proposed a conceptual upgrade that consists in replacing the classical concept of time of permanence in a given region of space, by the quantum concept of \textit{total availability} in lending itself to the creation of a spatial localization in that region. In the same way, we have proposed to replace the classical concept of time-delay by the quantum concept of \textit{total availability shift}. And to do so, we have reinterpreted \textit{quantum probabilities} as a gauge of the \textit{availability} of quantum entities in being part of an experience (or experiment) and produce the expected result.   

In the present analysis we have only considered the total availability of a quantum entity in creating a spatial localization in our three-dimensional physical space. However, nothing prevents us from extending the notion and consider more general spaces in which a quantum entity would actualize its presence. In general terms, these other spaces are characterizable in terms of projections operators, defining specific subspaces of the entity's Hilbert space ${\cal H}$. 

For instance, let $P_{\Delta E} = \int_{E_1}^{E_2} dE|E\rangle\langle E|$ be the projection operator into the set of states having their energy in the interval $\Delta E=[E_1,E_2]$. Then, we may ask what is the total availability of a free evolving quantum entity in manifesting its energy in $\Delta E$. If its initial state at time $t=0$ is $|\varphi\rangle$, then its total availability in $\Delta E$ is given by
\begin{equation}
\label{energy interval}
T_\varphi(\Delta E) = \int_{-\infty}^\infty dt\, \|P_{\Delta E}e^{-\frac{i}{\hbar}H_0t}\varphi\|^2 = \int_{-\infty}^\infty dt\, \|P_{\Delta E}\varphi\|^2.
\end{equation}

Clearly, (\ref{energy interval}) is infinite if $P_{\Delta E}|\varphi\rangle\neq 0$, and zero otherwise. This is because free evolution conserves energy, so that if at a given moment the entity has actualized its energy inside $\Delta E$, so it will be for all times. Seemingly, if at a given moment the entity has actualized its energy outside of $\Delta E$, so it will be for all times. In that sense, we can say that the free evolving quantum entity behaves locally in the one-dimensional energy-space (this is of course also the case for an interacting entity, provided the interaction is time independent).

As a last example, let us consider the concept of total availability in a relational sense. Instead of asking what is the total availability of a quantum entity in (actualizing its potential in) a given region of a given space, we may ask what is its total availability towards another quantum entity. Let $|\varphi_t\rangle$ and $|\chi_t\rangle$ be the states describing two quantum entities at time $t$, respectively. Then, we define the total availability of the first entity towards the second one (or of the second one towards the first, the relation being symmetric) by the integral
\begin{equation}
\label{reciprocal availability}
T_{\varphi,\chi} =\int_{-\infty}^\infty dt\, |\langle \chi_t |\varphi_t\rangle|^2.
\end{equation}
If the two entities evolve according to the same dynamics, i.e., $|\varphi_t\rangle = e^{-\frac{i}{\hbar}Ht} |\varphi\rangle$ and $|\chi_t\rangle = e^{-\frac{i}{\hbar}Ht} |\chi\rangle$, then $\langle \chi_t |\varphi_t\rangle = \langle \chi |\varphi\rangle$, so that $T_{\varphi,\chi}=0$ if $\langle \chi |\varphi\rangle =0$ (i.e., if at some instant the two entities are in orthogonal states), and $T_{\varphi,\chi}=\infty$, otherwise. In other terms, if the two entities, at some moment, have a certain degree of reciprocal availability, and if they evolve according to the same dynamics, they will remain mutually available in the course of their evolution, so that their total mutual availability is infinite. 

A typical example is two one-dimensional free evolving entities, such that $P_+|\varphi\rangle =|\varphi\rangle$ and $P_-|\chi\rangle =|\chi\rangle$, i.e., that are localized in the subspaces of positive and negative momentum, respectively (in classical terms we would say that they propagate in space  in opposite directions, but this wouldn't be a correct statement for quantum non-spatial entities). This means that, in a restrictive one-dimensional universe, we need to be oriented exactly in the same direction to be mutually available.    

On the other hand, if the two quantum entities do not evolve according to the same evolution, then we can expect $T_{\varphi,\chi}$ to also take finite values. Let us consider the case where $|\varphi_t\rangle = e^{-\frac{i}{\hbar}H_0t} |\varphi\rangle$ is free evolving, whereas $|\chi_t\rangle = e^{-\frac{i}{\hbar}E_0t} |\chi_{E_0}\rangle$ is the ground state of a potential $V$, of energy $E_0<0$. Then, 
\begin{equation}
\label{availability to a bound state}
T_{\varphi,\chi} =\int_{-\infty}^\infty dt\, |\langle \chi_{E_0}| e^{-\frac{i}{\hbar}H_0t}|\varphi\rangle|^2.
\end{equation}
If $V$ is for instance the multiplication operator by the one-dimensional delta-function $\lambda\delta(x)$, $\lambda <0$, and we assume $P_+|\varphi\rangle =|\varphi\rangle$, then a simple calculation shows that
\begin{equation}
\label{availability to a bound state2}
T_{\varphi,\chi} =\int_0^\infty dE\, \gamma(E,E_0) |\varphi(E)|^2,
\end{equation}
where 
\begin{equation}
\label{gamma function}
\gamma(E,E_0)=\frac{1}{v}\left(\frac{|\lambda|^3 m}{\hbar^2}\frac{1}{E+|E_0|}\right).
\end{equation}
Taking the monoenergetic limit (\ref{monoenergetic limit}) of a quantum free evolving entity that remains perfectly localized in energy, we find that
\begin{equation}
\label{availability to a bound state3}
T_{\varphi,\chi} \to \gamma(E,E_0),
\end{equation}
showing that the closer is the energy of the free evolving particle to the threshold of zero-energy (and therefore to the bound-state energy $E_0$), the higher is its total availability with respect to the (well localizable) bounded entity. 

We conclude by quoting de Ronde~\cite{Ron}: ``To a great extent every new theory that has been developed, from Aristotelian mechanics
to general relativity, has been grounded in new concepts. \textit{The physicist should be a creator of physical concepts}. Concepts which, within a theory, make possible to grasp certain character of nature. This however should not be regarded as some kind of solipsism, \textit{it is not only the description shaping reality but also reality hitting our descriptions}. It is through this interaction, namely, our descriptions and the
experimental observation that we create and discover a certain character of the being. It is in this way
that we can develop that which we consider to be \textit{reality}.''

In this note we have fully embraced the above program, and we hope that our modest proposal will not be knocked too vehemently by reality. Our suggestion is to take seriously the concept of \textit{availability}, that was introduced by Aerts in his creation-discovery view~\cite{Aer, Aer2}, by linking it directly to the quantum probability. And in doing so, we have provided what we think is a natural generalization of the classical concept of permanence. 

Permanence, in general terms, is total availability, but total availability doesn't reduce to permanence. For instance, we humans know that being available to another person means to allow ourselves to be placed in the other's shoes, i.e., to assume, for a certain time, the condition (the state) of the other. But we also know that, being available is not only a question of how much time we dedicate to the other, but as well the quality of it, that is, the efficiency and effectiveness with which we are open to experience something from the other's point of view. In this context, it is easy to understand why the concept of total availability is an upgrade of the concept of time of permanence: we can remain for a long time with a person and yet show a reduced total availability. But also, we can stay a very short time with a person and nevertheless manifest a high total degree of availability.

\end{document}